\documentclass[letter]{aa} 

\usepackage{graphicx}
\usepackage{txfonts}
\def\ltsima{$\; \buildrel < \over \sim \;$}
\def\simlt{\lower.5ex\hbox{\ltsima}}
\def\gtsima{$\; \buildrel > \over \sim \;$}
\def\simgt{\lower.5ex\hbox{\gtsima}}
\def\cgs{{erg cm$^{-2}$ s$^{-1}$}}
\def\ergs{{erg s$^{-1}$}}
\def\cm2{{cm$^{-2}$}}

\def\lum{{$L_{\rm 2-10}$}}
\def\lums{{$L_{\rm 0.5-2}$}}
\def\p1{{Paper I}}

\def\xmm{{\em XMM--Newton}}

\def\xmm{{\em XMM--Newton}}

\def\nh{{N$_{\rm H}$}}

\def\wise{{\em WISE}}

\def\f14{{10$^{-14}$}}
\def\f13{{10$^{-13}$}}
\def\f12{{10$^{-12}$}}
\def\f11{{10$^{-11}$}}

\def\4u{{4U~1344$-$60}}

\def\feka{{Fe K$\alpha$}}

\def\lmir{{$L_{\rm 5.8\mu m}$}}

\def\lir{{$L_{\rm 8-1000\mu m}$}}
\def\lbol{{$L_{\rm Bol}$}}
\def\w18{{W1835$+$4355}}
\def\mic{{$\mu$m}}
\def\spitzer{{\it Spitzer}}
\def\herschel{{\it Herschel}}
\def\msun{{$M_{\rm \odot}$}}

\begin{document}

   \title{The hidden quasar nucleus of a WISE-selected, hyperluminous,  dust-obscured  galaxy at $z$ $\sim$ 2.3}
           

   \author{E. Piconcelli\inst{1}, C. Vignali\inst{2,3}, S.~Bianchi\inst{4}, L.~Zappacosta\inst{1},  J.~Fritz\inst{5} , G. Lanzuisi\inst{3}, G.~Miniutti \inst{6}, A. Bongiorno\inst{1}, C. Feruglio\inst{7}, F. Fiore\inst{1},  R. Maiolino\inst{8}}

 \titlerunning{X-ray observation of a Hot DOG at $z$=2.3}\authorrunning{E.~Piconcelli et al.}


       \institute{Osservatorio Astronomico di Roma (INAF), Via Frascati 33, I--00040
  Monte Porzio Catone (Roma), Italy \and  
  Dipartimento di Fisica e Astronomia, Universit\`a  di Bologna, Viale Berti Pichat 6/2, I-40127 Bologna, Italy \and
  Osservatorio Astronomico di Bologna (INAF),  Via Ranzani 1, I--40127 Bologna, Italy \and 
Dipartimento di Matematica e Fisica, Universit\`a degli Studi Roma Tre, Via della Vasca Navale 84, I--00146 Roma, Italy  \and 
Sterrenkundig Observatorium Universiteit Gent, Krijgslaan 281, S9, 9000, Gent, Belgium \and
Centro de Astrobiologia (CSIC-INTA), Dep. de Astrof\'{i}sica, ESAC, PO Box 78, E-28691, Villanueva de la Ca\~nada, Madrid, Spain \and
Institute de Radioastronomie Millimetrique, 300 Rue de la Piscine, F-38406 St. Martin d’Heres, Grenoble, France \and
Cavendish Laboratory, University of Cambridge, 19 J. J. Thomson Avenue, Cambridge CB3 0HE , UK 
}  

   \date{Received}

 
  \abstract
    {We present the first X-ray spectrum of a hot dust-obscured galaxy (DOG),  namely W1835$+$4355 at $z$ $\sim$ 2.3.
   Hot DOGs represent a very rare population of  hyperluminous ($\geq$ 10$^{47}$ \ergs), dust-enshrouded objects at $z$ $\geq$ 2 
   recently discovered in the {\it WISE} All Sky Survey.  
The 40 ks \xmm\ spectrum   reveals  a  continuum as  flat ($\Gamma$ $\sim$ 0.8) as typically seen in heavily obscured AGN. 
This, along with  the presence of strong \feka\ emission, clearly 
suggests a reflection-dominated spectrum due to  Compton-thick absorption.
In this scenario, the observed luminosity of \lum\ $\sim$ 2 $\times$ 10$^{44}$ \ergs\ is a fraction ($<$10\%) of the
intrinsic one, which is estimated to be
\simgt5 $\times$ 10$^{45}$ \ergs\ by using several proxies. The \herschel\ data allow us to constrain the SED up to  
the sub-mm band, providing a
reliable estimate of the quasar contribution ($\sim$ 75\%) to the IR luminosity as well as the amount of star formation ($\sim$ 2100 M$_{\odot}$ yr$^{-1}$).
Our results thus provide additional pieces of evidence that associate Hot DOGs with an exceptionally dusty phase  during which
 luminous quasars and massive galaxies co-evolve and  a very efficient and powerful AGN-driven feedback mechanism is predicted by models.}
   
   \maketitle
%

\section{Introduction}
 
 The detection and the study of  the most luminous quasars (\lbol\ $\gg$ 10$^{46}$ \ergs) at $2<z<3$, where   their  number 
density reaches a peak,  is  essential
 to understanding the major formation events over the  supermassive black hole (SMBH) assembly history as well as to probing the  co-evolution of host  galaxies and their central SMBH at its extremes.
For the luminous systems, models predict a merger-induced evolutionary sequence with 
an initial heavily dust reddened phase associated with strong and obscured  SMBH growth 
and  star formation (Silk \& Rees 1998; Hopkins et al. 2008; Fabian 2012).
There is growing evidence that  the   red phase coincides with  the  blow-out (i.e., AGN-driven feedback-dominated; Faucher-Giguere \& Quataert 2012) phase,  after which the quasar eventually evolves into a  blue, optically bright source (Banerji et al. 2012; Glikman  et al. 2012; Brusa et al. 2015).

Red luminous quasars are thus ideal test-beds for this AGN-driven feedback scenario. 
{\it Spitzer} observations allowed the discovery of a population of dust-obscured galaxies (DOGs hereafter)
 with extreme mid-IR/optical ratios ($F_{24~\mu {\rm m}}$/$F_{\rm R}$ $>$1000) and a flux density 
 at 24$\mu$m $F_{24~\mu {\rm m}}$ \simgt~1 mJy at  $1<z<3$  (Dey et al. 2008).
The AGN nature of DOGs is suggested by their  mid-IR power-law spectrum and has been definitively 
confirmed via X-ray observations (Fiore et al. 2008; Lanzuisi et al. 2009, L09), which   found a 
 predominance of type-2 quasars  (\lum\ $>$  10$^{44}$ \ergs\ and \nh\ $>$ 10$^{22}$ \cm2) among them.
\begin{figure*}
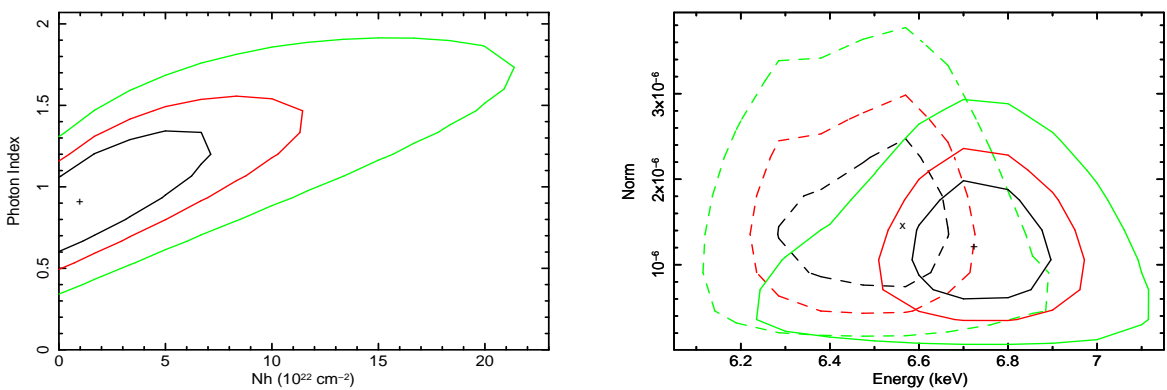

\begin{center}
\includegraphics[width=5cm,height=7.5cm,angle=-90]{fig1a.ps}
\hspace{0.5cm}\includegraphics[width=5cm,height=7.5cm,angle=-90]{fig1b.ps}
\caption{{\bf(a)}--{\it Left:} 
Confidence contour plot for the photon index against the column density using  an absorbed PL model. 
{\bf(b)}--{\it Right:} Confidence contour plot showing the normalization against energy of the \feka\ emission line for the PN (solid lines) and MOS
 (dashed lines) spectrum. 
 The contours are at 68\%, 90\%, and 99\% confidence levels for two interesting
parameters.
}
\label{spettri}
\end{center}
\end{figure*}
The study of rare, very luminous DOGs is now taking advantage of the
recent sensitive,  wide-area surveys such as the all-sky  \wise\ survey (Wright et al. 2010). In particular,
by selecting "W1W2-dropout"  objects, i.e.,  bright at 22 or 12 \mic\ but  very faint at 4.6 and 3.4  \mic\ (\wise\ W4, W3, W2, and W1 bands, respectively),   
Eisenhardt et al. (2012)  identified a population of $\sim$ 1000 DOGs
 which stand out for being hyperluminous (\lbol\ $\sim$ 10$^{47}$ \ergs) at $z$ $\sim$ 2--3.
 They were dubbed Hot DOGs because their spectral energy distributions (SED) typically exhibit  
 a mid-IR/sub-mm emission ratio higher than other galaxy types, suggesting a large contribution  from hot dust (Wu et al. 2012, 2014; Jones et al. 2014).
  Optical follow-up reveals narrow emission lines in most of their spectra.  
   Hot DOGs show no  clear-cut sign of gravitational lensing and, therefore, are truly hyperluminous.
   Since they likely represent a short-lived crucial  phase in the evolution of the  most powerful  AGN in the universe,
 probing the nuclear properties of such recently-discovered extreme  objects can clearly benefit  our understanding of overall AGN-galaxy co-evolution. 
 
 This  is our aim and here we report the first X-ray spectrum of a Hot DOG (namely WISE J183533.71$+$435549.1,
 hereafter \w18, at $z$ = 2.298, e.g., Wu et al. 2012) obtained by  \xmm.  Throughout this Letter  we assume
 $H_0$ = 70 km s$^{-1}$ Mpc$^{-1}$,
$\Omega_\Lambda$ = 0.73, and  $\Omega_M$ = 0.27.

\section{X-ray observation and data reduction}

\w18\ was observed with \xmm~in revolution 2508 (18 August 2013)
 for about 42 ks (Obs. ID. 0720610101).
The EPIC observations were
performed with the PN and MOS cameras operating in
Full-Window mode and with the Thin and Medium  filter applied, respectively.
The extraction of science products from the observation data files followed standard 
procedures using the \xmm\ Science Analysis System SAS v13.5.
X-ray events corresponding to
patterns 0--4(0--12) for the PN(MOS)~camera were selected. 
The event lists were filtered to ignore periods of high background flaring activity
by selecting good time intervals with  count rate $<$ 0.4(0.35) counts s$^{-1}$
using  single event PN(MOS) light curves at $E$ $>$ 10 keV.
For the PN, the source photons were extracted with a  circular region of 26 arsec radius, with the background being derived
from a source-free 57 arcsec radius region on the same chip.
MOS1 and MOS2 source(background) spectra were extracted using a circular region of radius 14(45) arcsec.
The useful exposure times for spectroscopy was 35.7 and 41.2 ks for PN and  MOS, respectively.
Finally, we created  a combined MOS spectrum and response matrix using {\it addascaspec}.

\section{Results}
\label{s:results}

\w18\ is well detected in both EPIC cameras at R.A. (J2000) = 18:35:33.76 and Dec. (J2000) = $+$43:55:48.6, with a 0.3-10 keV net count rate of 
 3.4$\pm$0.4,  1.4$\pm$0.3, and 1.3$\pm$0.2 $\times$ 10$^{-3}$ counts s$^{-1}$ in PN, MOS1 and MOS2, respectively.  
Spectral analysis was carried out using the modified Cash statistic ($C$-stat; Cash 1979) 
provided in XSPEC.
The PN(MOS) spectrum was rebinned with a minimum  of 10(7) counts bin$^{-1}$ so that MOS and PN spectra 
contained a comparable number of spectral bins.
 In the following, errors correspond to the 90\% (1.6$\sigma$) confidence level for one interesting parameter, i.e., $\Delta$$C$ = 2.71.
A Galactic column density of \nh\ = 5.2 $\times$ 10$^{20}$ \cm2\ (Kalberla et al. 2005) was applied to all spectral models.
A power law (PL)  model yielded statistically consistent  spectral parameters with a slope
 $\Gamma$ $\sim$ 0.8 and a normalization $\sim$ 1.3 $\times$ 10$^{-3}$ photons/keV/cm$^2$/s for separate fits to the PN and MOS spectra. 
 The PN and MOS data were then  fitted simultaneously in the 0.5--8 keV band.
As a very flat spectral index is a typical indication of an absorbed continuum, we added  an intrinsic absorption component to the model. 
The best-fit value of  $\Gamma$ remained, however, basically the same  and we placed an upper limit of \nh\ $<$ 8 $\times$ 10$^{22}$ \cm2\
for the column density of the obscuring material ($C$-stat/dof = 42/30). 
Figure 1a shows the confidence contour plot of the column density  of the cold absorber against the slope of the PL.
If a canonical PL with $\Gamma$ = 1.8  were present, this would imply a \nh\ \simgt\ 10$^{23}$ \cm2\ of the obscuring material.
Interestingly, a positive excess in the fit residuals was present around 2 keV (i.e., $\sim$ 6.4 keV rest frame), 
so a Gaussian emission component was added. This model provided a good description of the X-ray spectrum, with $C$-stat/dof = 22/29.
 The rest-frame energy of the line was 6.63$\pm$0.09 keV,  with a normalization of 1.0$\pm$0.3 $\times$  10$^{-6}$ photons/keV/cm$^2$/s.
 We measured a rest-frame equivalent width (EW) of the line EW = 2.2$^{+1.1}_{-0.9}$ keV. 
 The presence of such a strong emission feature  is very interesting since
 reflection-dominated spectra of Compton-thick AGN are indeed characterized by a very flat photon index and prominent
  (i.e., EW \simgt\ 0.8 keV) iron K lines from neutral and ionized species (e.g., Matt et al. 2004; Nandra \& Iwasawa 2007; Tilak et al. 2008). 
 We  therefore performed an additional investigation on this spectral feature as it was tempting to associate the 
 X-ray emission of \w18\ with a completely buried quasar scenario.
We first checked  for the presence of the line in the PN and MOS spectrum separately. The significance of the line detection
 was at $>$ 99\% confidence level for  two parameters (i.e., energy and normalization) in both datasets, 
 with  the centroid being at 6.73 $\pm$ 0.19 keV (PN) and 6.55$^{+0.13}_{-0.29}$ keV (MOS), i.e., consistent at 68\% confidence level (Fig. 1b).
We then  inspected  the properties of the \feka\ line using {\it unbinned} (i.e., with a $\Delta$E  $\sim$ 50 eV) PN and MOS spectra in the 1.6-2.5 keV band. 
In both cases, the underlying continuum was fitted by a PL with $\Gamma$ = 0.9. The excess was confirmed in both datasets.
 In particular,  it can be modeled by two narrow emission lines  in the PN spectrum
  (which is the dataset with the best statistics, i.e., with a count PN/MOS ratio of $\sim$1.3 in this energy interval). 
  We fixed the energy of these lines at 6.4 and 6.67 keV to model the neutral Fe and FeXXV emission, respectively. 
  As  a further consistency check, we tied the redshift of the lines and  left it free to vary. 
  The best-fit value of  $z$ was found to be 2.28$^{+0.09}_{-0.08}$, i.e., consistent with the $z$ = 2.298 reported by Wu et al. (2012). 
  This suggested a likely blend of the two emission lines in the binned spectrum. 
 The excess in the unbinned MOS spectrum can be fitted by a single line with a  centroid at 6.54$\pm$0.09 keV.
  
\begin{figure}[]
\begin{center}
\includegraphics[width=5cm,height=8.cm,angle=-90]{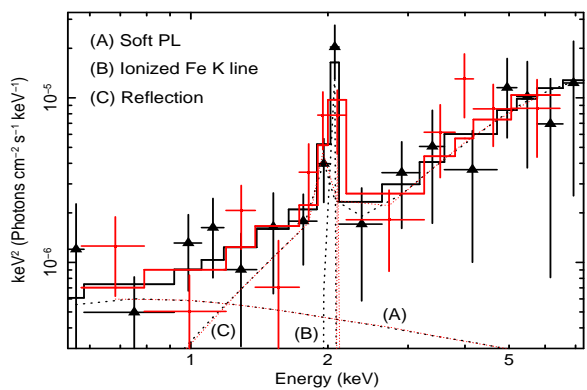}
\caption{The  reflection-dominated model fitted to the unfolded PN (triangles) and MOS  (circles) spectrum of \w18. 
The individual spectral components (i.e., soft PL, ionized Fe emission line, and neutral reflection) are also shown as dotted lines.
}
\label{cont}
\end{center}
\end{figure}
  The combination of  intense  \feka\ emission superimposed on a flat continuum in \w18\ led us to 
  consider a pure reflection scenario for the X-ray emission from the AGN in this Hot DOG. To this end, 
  we used the PEXMON model in XSPEC (Nandra et al. 2007) which includes both neutral Compton reflection  
  and fluorescence neutral Fe emission lines. 
   Given the above results,  we also added a  narrow Gaussian line to model the ionized Fe emission line. 
    Assuming this spectral model, we obtained $C$-stat/dof = 24/29.
  The additional  line was centered at 6.65$^{+0.18}_{-0.11}$ keV, while the best-fit value of photon index 
  of the illuminating continuum was $\Gamma$ $\sim$ 2.4, i.e.,  steeper than the mean  quasar index (i.e., 1.8--1.9; Piconcelli et al. 2005). 
  The X-ray spectrum of heavily obscured AGN is characterized by the presence of a "soft excess", which typically extends up to $\sim$ 3 keV, and is
  due to line-dominated emission from large-scale photoionized gas and, in the case of star-forming galaxies, from hot, collisionally-ionized gas in starbursts.
   In CCD-like resolution and/or low quality spectra, a steep PL component ($\Gamma$ $\approx$ 2.3--2.7, e.g., Turner et al. 1997; Comastri et al. 2010) 
   is usually applied to account for this soft X-ray emission. Accordingly, we then added a PL with $\Gamma$ fixed to 2.5 to the previous reflection model.
    This model yielded an excellent description to the data (see Fig. 2)  without statistically significant unfitted features ($C$-stat/dof = 20/28), 
    and a best-fit value $\Gamma$ = 2.0$^{+0.5}_{-0.7}$ for the intrinsic continuum slope. 
 We measured a rest-frame  EW = 990$^{+760}_{-490}$ eV for the ionized Fe  line and a ratio between the normalization of the soft and intrinsic hard PL 
  $\sim$ 0.03, which is a typical value for heavily obscured AGN (Turner et al. 1997; Guainazzi et al. 2005).
  
This model gives a 2--10(0.5--2) keV  flux  of 2.11(0.28) $\times$ 10$^{-14}$ \cgs, 
which corresponds to an observed luminosity  \lum\ = 2.3 $\times$ 10$^{44}$ \ergs\ (\lums\ = 2.9 $\times$ 10$^{43}$ \ergs).

\section{SED of \w18}
\begin{table}
\caption{Photometry of \w18\ from near-IR to sub-mm.}
\label{table:1}
\centering    
\begin{tabular}{l c c c }
\hline\hline\\          
Instrument & Band & Flux Density (error) & Ref.\\
\hline                      
\spitzer\ IRAC1  &   3.6\mic\ & 51.5  (2.2)  $\mu$Jy & Wu12\\
\spitzer\ IRAC2  &   4.5\mic\ & 142.8 (3.0)  $\mu$Jy & Wu12\\
\wise\ W3        &  12\mic\   & 6790  (190)  $\mu$Jy & IRSA\\ 
\wise\ W4        &  22\mic\   & 24.6  (1.0)  mJy     & IRSA\\
\herschel\ PACS  &  70\mic\   & 40.9  (4.1)  mJy     & this work \\
\herschel\ PACS  & 160\mic\   & 79.3  (7.9)  mJy     & this work \\
\herschel\ SPIRE & 250\mic\   & 81.0  (8.1)  mJy     & this work \\
\herschel\ SPIRE & 350\mic\   & 72.0  (7.2)  mJy     & this work \\
\herschel\ SPIRE & 500\mic\   & 33.0  (3.3)  mJy     & this work \\
{\it SCUBA-2}    & 850\mic\   &  8.0  (1.5)  mJy     & J14\\ 
\hline\hline        
\end{tabular}
\tablefoot{References: Wu12: Wu et al. (2012); J14: Jones et al. (2014). 
Notes: \spitzer\ IRAC data are derived from warm-mission observations; 
\wise\ data are taken from the AllWISE source catalog at IRSA. 
Nominal W4 flux density has been corrected by a factor of 0.90 which applies 
for steeply rising mid-IR SEDs (see http://wise2.ipac.caltech.edu/docs/release/allsky/expsup/sec4\_4h.html). 
Errors of 10\% are reported for \herschel\ flux densities.}
\end{table}

\begin{figure}[b]
\begin{center}
\includegraphics[width=0.35\textwidth,angle=-90]{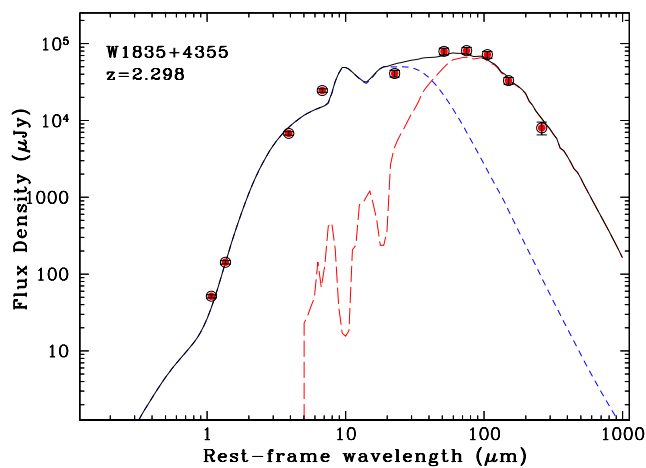}
\caption{Rest-frame SED of \w18\ (see Table~\ref{table:1} for the available 
photometry). The best-fitting SED (black continuous curve) derives from the 
contribution of an AGN (blue short-dashed line) and a star-forming component 
(red long-dashed curve); see text for details.}
\label{sed}
\end{center}
\end{figure}

To have a better understanding of \w18\ properties, we collected the available 
photometric information from literature and extracted flux densities from 
archival \herschel\ observations (see Table~\ref{table:1}). 
To this end, we used HIPE to extract the  70, 160, 250, 350, and 500 \mic\  source flux densities from 
the  level2 fits files of \w18\ using circular regions of size 12, 12, 
22, 30, and 42 arcsec, respectively; background was chosen from an annulus centered on the source 
(with an inner radius larger than the source extraction radius to avoid 
contamination by the source PSF tails), and proper aperture corrections 
were applied (as described in the PACS and SPIRE Data Reduction Guides).   
The source is detected from the observed mid-IR bands (where both \spitzer\ 
warm-mission data and AllWISE data are available) to the sub-mm band 
({\it SCUBA-2}). This broadband coverage allows us to provide a good modeling 
of the SED in terms of an AGN plus a starburst component using the fitting 
code originally developed by Fritz et al. (2006), recently updated by 
Feltre et al. (2012); see Vignali et al. (2011) 
for a recent application of this code. 
For the hot AGN component, the code uses an extended grid of 
{\it smooth} torus models with a flared disc geometry (Fritz et al. 2006), 
which typically provides a good description of the AGN emission even in cases 
where only sparse photometric datapoints are available (see Feltre et al. 2012). 
For the far-IR emission, an additional component due to colder 
diffuse light, heated by star formation processes, is included in the fitting 
procedure. The fit to this emission is carried out using templates of known 
starburst galaxies. 
As shown in Fig.~\ref{sed}, the AGN component (short-dashed line) is able to 
reproduce the emission at short wavelengths reasonably well, while in the 
far-IR, additional emission, linked to star formation activity and here 
parameterized with the Arp220 template, is needed. 
The adopted Arp220 template is semi-empirical, calculated with the
GRASIL code (Silva et al. 1998). We also used several other templates
of local starbursts to try to account for the far-IR emission of \w18, but
none of them produced acceptable results. Fitting the far-IR starburst-related
emission with a modified blackbody model (fixing the emissivity index  $\beta$ = 2)
provided an acceptable fit  with T$_{dust}=39.9\pm1.2$~K and
M$_{dust}=1.3\pm0.2\times10^{9}$ M$_\odot$, and a far-IR luminosity consistent with
that inferred by our best-fitting model.
The obscured AGN SED model alone can naturally account for the optical emission. 
This result can, in principle, be justified by the very high AGN luminosity; however, it can be partially due to the poor sampling of the rest-frame optical/near-IR SED.
We can reproduce the data equally well using the same Arp220 template in  the near-IR/optical band.
In this case, a torus model with a slightly higher optical depth is required since part
of the optical emission, originally ascribed to the AGN, is provided by stars. 
We therefore cannot exclude a stellar contribution to the optical emission, but it would not significantly change any of the derived quantities.
From the best-fitting SED (Fig.~\ref{sed}) we can 
calculate both the \lbol\ related to accretion processes and 
the star formation rate (SFR). 
The AGN bolometric luminosity is $\approx4.8\times10^{47}$ \ergs, with 
a quasar contribution to the total \lir\ of $\sim$ 75\%. 
To estimate the SFR, we converted the 
rest-frame 8--1000 \mic\ luminosity due to star formation (\lir\ $\approx4.6\times10^{46}$ \ergs) in the 
best-fitting SED using the Kennicutt et al. (1998) formula and obtained  a SFR $\approx2100$~\msun\ yr$^{-1}$.

\section{Discussion}
The \xmm\ observation presented here provides the first X-ray spectrum of a Hot DOG and highlights  
the potential offered by present-day X-ray telescopes 
to probe the nuclear environment of such hyperluminous high-$z$ systems.
As expected on the basis of the  "W1W2-dropout"  selection criterion, which  preferentially 
picks up highly reddened objects, the X-ray continuum emission of \w18\ turns out to be very obscured.
An absorbed PL model indeed gives a flat   $\Gamma$ = 0.8, 
 indicative of  \nh\  $\gg$   10$^{23}$  \cm2.
The  large EW \feka\ line observed  at  $\sim$6.6 keV (rest frame) can be thus
interpreted as a signature of reflection-dominated spectrum due to a Compton-thick 
 absorber blocking the direct continuum emission along the line of sight to the nucleus of \w18.
 The presence of a typical soft excess PL component possibly suggests
that the X-ray absorber allows some continuum emission to
leak out. 
Previous 10--20 ks   PN observations of Hot DOGs  were unable to detect an X-ray source in two out 
of  three cases, with the only faintly detected source (W1814+3412 at $z$ $\sim$ 2.4) 
showing  an X-ray flux of  5 $\times$ 10$^{-15}$ \cgs\ (Stern et al. 2014), which further
 supports the presence of buried AGN in these galaxies.
In this scenario, the observed hard X-ray luminosity  of  \w18\ ($\approx$ 2 $\times$ 10$^{44}$ \ergs)
  must be considered only a fraction of the intrinsic one. The ratio observed/intrinsic \lum\ typically ranges
  from 0.05 to 0.012  for \nh\  values in the range 10$^{24}$  to  10$^{25}$\cm2\ 
   (Brightman \& Nandra 2011; Singh et al. 2012).
We are thus able to estimate a lower limit to the  intrinsic \lum\   of $\approx$ 4.5  $\times$ 10$^{45}$ \ergs\ for \w18.
Additional support for this range of \lum\ is given by  the empirical mid-IR--X-ray luminosity relations.
Specifically, we use the relation  obtained by L09 for a large sample of DOGs with good X-ray spectral information.
The \lmir\ derived from the SED is  $\approx$ 9 $\times$ 10$^{46}$ \ergs.
Accordingly, in the \lmir\ vs observed \lum\ plane shown in Fig. 6b of L09, the values
 for \w18\ suggest a \nh\ close to 10$^{24}$ \cm2.
The observed \lmir\ implies a \lum\ $\approx$ 4 $\times$ 10$^{45}$ \ergs, i.e., consistent with the X-ray--based estimate. 
Furthermore,  the AGN \lbol\ inferred in Sect. 4 translates into a \lum\ $\approx$ 5  $\times$ 10$^{45}$ \ergs\ 
once a bolometric correction of $\sim$ 100 is applied (Hopkins et al. 2007).
We note, however,  that such extreme luminosities are basically not sampled by these relations
 involving \lum\ and, therefore, these estimates  should only be regarded as indicative.

Our results lend support to Hot DOGs as tracers of an exceptional,   rapid, and 
 dusty phase of quasar/host galaxy co-evolution. The X-ray spectrum  and the SED of \w18\  indicate 
 that it is  indeed associated with the very luminous tip  of both AGN and galaxy populations.
  The \herschel\ data allow us to constrain the emission from cold dust heated by star formation 
  and infer that  the host galaxy is rapidly star forming at a rate of  thousands of \msun\ yr$^{-1}$,
 i.e., similar to sub-mm galaxies (SMGs) at $z \geq$ 2  (Hayward 2013). 
  Assuming a  gas-to-dust ratio of $\approx$ 50 as found in SMGs (Kov{\'a}cs et al. 2006) and
   M$_{dust}$ = 1.3 $\times10^{9}$ M$_\odot$, the measured SFR can be sustained only for $\sim$ 3 $\times$  10$^{7}$  yrs 
  unless additional gas is provided.

Models and simulations predict that the combination of  red phase and huge luminosity  offers the opportunity of  catching  the  AGN-driven feedback process  at a maximum level of efficiency (Hopkins et al. 2008),  
 and strong indications  of feedback-related features, i.e., $\sim$100 kpc extended Ly$\alpha$ blobs around some Hot DOGs, have indeed been reported (Bridge et al. 2013).   Hot DOGs are thus ideal targets for millimeter interferometry to search for massive, galaxy-wide molecular outflows.
 Additional, well-designed X-ray observations of Hot DOGs will obtain crucial information about  the poorly-studied nuclear environment of  the most extreme AGN in the universe.
Specifically,  spectra above 10 keV can provide an invaluable, absorption-free view of  their continuum source, responsible  for their   high luminosity.

\begin{acknowledgements}
 We thank the anonymous referee for helpful comments.
 EP and AB acknowledge financial support from INAF under the contract PRIN-INAF-2012.
\end{acknowledgements}


\end{document}